%% file: grebel.tex
\documentclass[multphys,vecphys]{svmult}

\usepackage{makeidx}         
\usepackage{graphicx}        
\usepackage{multicol}        
\usepackage[bottom]{footmisc}

\makeindex             

\begin{document}

\title*{Local Group(s)}
\author{Eva K.\ Grebel\inst{1}
}
\institute{Astronomical Institute of the University of Basel,
Department of Physics and Astronomy, Venusstrasse 7, CH-4102 Binningen,
Switzerland
\texttt{grebel@astro.unibas.ch}
}
%
%
\maketitle

\begin{abstract}
The properties of the galaxies of the Local Group are reviewed,
followed by a brief discussion of nearby groups.  The
galaxy groups in our vicinity -- the M81 group, the Cen\,A group, and
the IC\,342/Maffei group --  are in many respects
Local Group analogs:  Their luminosity functions, galaxy content, 
fractional galaxy type distribution, crossing times, masses, and
zero-velocity surface radii are similar to those of the Local Group.
Also, the nearby groups usually consist of two subgroups, some of
which approach each other and may ultimately merge to form a fossil
group.  These poor groups contrast with the less evolved, loose
and extended galaxy ``clouds'' such as the Scl group and the CVn\,I
cloud. These are characterized by long crossing times, are
dominated by gas-rich, late-type galaxies, and lack gas-deficient, low
luminosity early-type dwarfs.  These clouds may be groups still in
formation.  The local Hubble flow derived from the clouds and groups
is very cold.   
\end{abstract}

\section{Why are galaxy groups interesting?}
\label{sec:1}

The most conspicuous gatherings of mass and luminous matter in the
Universe are galaxy clusters.  However, the majority of nearby
galaxies -- about 85\% (\cite{Tully87, Kara05}) -- is observed to be
located outside of clusters and can be found mainly in galaxy groups.
While galaxy clusters reveal the location of the highest concentration
of visible and dark matter, these less massive galaxy agglomerations
trace the distribution of the filaments of the cosmic web and hence
the extended distribution of dark matter in less dense regions.  In
terms of luminous matter, the fraction of mass locked up in stars
increases with decreasing group size (\cite{Eke05}).  Most of the
stellar mass is in groups of the size of our Local Group, whereas
massive clusters only contain 2\% of the stellar mass in the Universe
(\cite{Eke05}).

Galaxy groups come in a variety of different morphologies, shapes, and
sizes, including, for instance, seemingly unbound ``clouds'' and
``spurs'' (\cite{Tully87}), loose groups, poor groups, compact groups,
rich groups, etc.  Precisely what constitutes a group is a question of
definition and depends primarily on how many galaxies down to a
certain limiting magnitude are found within a certain volume.  Common
usage often considers an agglomeration of some 30 galaxies within a
radius of one to two Mpc a galaxy group, but this broad definition is
usually adjusted to the practical needs of the actual application; for
instance, when searching for groups in redshift surveys (e.g.,
\cite{Lee04, Eke04}).  

The different types of richness and compactness of groups permit us to
study galaxy evolution and environmental effects in lower-density
regions, and to contrast the results with the properties and the
evolutionary state of galaxies observed in the field or in rich
clusters.  The degree of clustering of groups increases with
increasing group luminosity (\cite{Padilla04}).  Within groups, the
earlier-type galaxies are more strongly clustered and tend to lie
closer to the centers of the groups (\cite{Girardi03}), similar to the
well-known morphology-density relation in galaxy clusters
(\cite{Dressler80, Binggeli87}). The final stage of evolution within
groups may be represented by fossil groups, which are dominated in
light and mass by a large central elliptical galaxy, presumably the
end product of major mergers (\cite{Jones03, D'Onghia05}).  The
different types of groups appear to epitomize different stages from
early to advanced structure formation. 

Members of galaxy groups and ensembles of groups are kinematic tracers
that can be useful for mass determinations and that help to uncover
the local properties of large-scale galaxy flows, such as the local
Hubble flow.  It has even be suggested that they may reveal the
effects of dark energy on small scales ($>7$ Mpc; e.g., \cite{Kara02a,
Kara03a, Maccio05}).  

The title of this contribution as assigned to me can be interpreted in
two ways:  Apart from the Local Group, one may understand it to cover
galaxy groups that are local, i.e., nearby, or one can interpret it as
referring to galaxy groups that are counterparts of the Local Group.
In fact, the nearby galaxy groups often combine these two properties.
Thus in this review I will concentrate mainly on the Local Group and
nearby Local Group analogs.

Our own Milky Way is a member of a poor groups of galaxies, the Local
Group.  The nearest neighboring group, the Sextans-Antlia group, is a
an even poorer group consisting only of dwarf galaxies -- sometimes
also referred to as a dwarf group \cite{vdB99a}, \cite{Tully02}.
Additional poor groups are located in our vicinity.  In a number of
ways we live in a fairly average extragalactic neighborhood.

\section{The Local Group}

The Local Group is the home to two large spiral galaxies, our Milky
Way and M31.  The Local Group is the only place where we can study the
ages, chemistry, star formation history, and kinematics of a range of
different galaxies in exceptional detail based on their resolved
stellar populations.  Here we can truly connect stellar and
extragalactic astrophysics.  We can compare our observational findings
with the predictions of cosmological models or, vice versa, test those
predictions through targeted observations.  Owing to our ability to
resolve and study even the oldest populations and very faint stars in
the closest galaxies and in our own Milky Way, we can conduct
near-field cosmology (\cite{Freeman02}), uncovering their detailed
evolutionary histories based on their resolved fossil stellar record.

\subsection{The Local Group census and galaxy distribution}

In spite of their proximity and in spite of the large efforts invested
into studying them over the past decades, the galaxies of the Local
Group continue to provide surprises.  As has also been found in other
groups, the number of faint galaxies in the Local Group lies below
theoretical expectations by about two orders of magnitude, a
deficiency also known as the missing satellite problem or the
cosmological substructure problem (\cite{Klypin99, Moore99}).  A
number of solutions to solve this problem have been proposed, but none
has proved fully satisfactory until now.  This problem remains one of
the key questions in structure formation in standard cold dark matter
(CDM) models.  Nonetheless, considerable progress has been made in
recent years in identifying new Local Group members.  Within a
zero-velocity radius of approximately 1 Mpc (\cite{Kara02a}) the
current Local Group census comprises at least 42 galaxies (including
the tidal streams of Sagittarius around the Milky Way and the great
stellar stream around M31, \cite{Ibata94, Ibata01}).    

The three most luminous Local Group members are its spiral galaxies,
which form two subgroups:  The M31 plus M33 subgroup and the Milky Way
subgroup.  In terms of mass and luminosity, the Milky Way and M31 are
the two dominant galaxies and may be similarly massive
(\cite{Evans00}).  Each of these two galaxies is surrounded by an
entourage of mainly low-mass, gas-deficient galaxies.  In close
proximity to M31, one compact elliptical and three dwarf elliptical
(dE) galaxies are found.  M31 is the only Local Group galaxy with dE
companions.  The remaining low-mass early-type dwarfs in the Local
Group are all dwarf spheroidal (dSph) galaxies, the least massive
(total masses estimated to be a few times $10^7$ M$_{\odot}$), least
luminous ($M_V > -14$) type of galaxy known.  (For a more detailed
description of the different galaxy types of the Local Group, see
\cite{vdB00, Grebel01, GGH03}.) The dSphs and the dEs are almost all
found within a 300 kpc radius around the two dominant spirals.  This
radius also roughly corresponds to the size of the dark matter density
profile of $\sim$ L$_{\star}$ galaxies (e.g., \cite{McKay02,
Prada03}). The gas-rich, late-type irregular and dwarf irregular
(dIrr) galaxies, in contrast, show a much less concentrated
distribution and and are the most frequent galaxy type at larger
distances from the spirals.  

This biased distribution is also apparent when considering the H\,{\sc
i} mass of dwarf and satellite galaxies as a function of distance to
the nearest principal galaxy:  The upper limits for the H\,{\sc i}
masses of dSphs are typically of the order of $10^5$ M$_{\odot}$ or
less, whereas dIrrs usually have H\,{\sc i} masses of at least $10^7$
M$_{\odot}$ (\cite{GGH03}).  In between, the so-called dIrr/dSph
transition-type galaxies are located.  These low-mass dwarf galaxies
share some of the properties of dIrrs (such as a measurable gas
content and recent star formation) and of dSphs (such as prominent old
populations and low luminosities).  Morphological segregation akin to
that within the Local Group is also observed in other groups and
indicates that environment plays an important role in shaping galaxy
properties (e.g., \cite{Einasto74}).  The distribution of the galaxies
in the Local Group is illustrated in, e.g., \cite{Grebel97, Grebel99}.  

More than half of the Local Group's galaxies are dSph galaxies.  At
present, we know of 22 dSphs in the Local Group.   Three of the M31
dSph companions were only detected and confirmed approximately seven
years ago (\cite{Armand98, Armand99, KK99, GreGuh99}).  Two more M31
dSph companions were discovered during the past two years in the Sloan
Digital Sky Survey (SDSS) and confirmed through follow-up observations
by \cite{Zucker04a, Harbeck05, Zucker06a}. They are among the least
luminous, lowest surface brightness dwarfs known.  Altogether eight
confirmed or likely dSph companions of M31 are now known.  Additional
faint dSphs may yet to be found, whereas other features (e.g.,
\cite{Morrison03, Zucker04b}) are likely part of the giant stellar
stream around M31 (\cite{Ibata04, Fardal06}).  Recent additions to the
Milky Way dSph census include three extremely low surface brightness
dwarfs also found the SDSS (\cite{Willman05, Belokurov06a,
Zucker06b}), which increases the number of Galactic early-type
companions to 12.  The new dSph Boo is the faintest galaxy known so
far with $M_V \sim -5.7$ (\cite{Belokurov06a}), and the new CVn dSph
seems to have the lowest surface brightness of any of these dwarfs
($\mu_V \sim 28$ mag arcsec$^{-2}$; \cite{Zucker06b}).  The recent
detections beg the question whether there is a lower-mass cut-off for
dwarf galaxies with luminous baryonic matter.  Finally, some seven
years ago a new {\em isolated} Local Group dSph was discovered by
\cite{Whiting99}, raising the number of these rare objects to two.
Isolated dSphs are of particular interest since they seem to defy
commonly held ideas about the creation of dSphs through, e.g., tidal
or ram pressure stripping by massive galaxies.  On the other hand,
since we do not know the orbits of these more distant dSphs, we cannot
exclude that they may once have had close encounters with the massive
galaxies.  

All recent galaxy discoveries added more objects to the faint end of
the Local Group's luminosity function.  It is generally believed that
the census is fairly complete for brighter galaxies (with the possible
exception of the zone of avoidance.)  Additional very low surface
brightness objects may be uncovered in the coming years using large
imaging sky surveys.  We may expect that nearby groups also host a
number of comparatively faint objects that have escaped detection so
far.  In spite of the recent impressive improvements in the census of
nearby galaxies (see, e.g., \cite{KKSG00, Kniazev04b} and references
therein) the numbers are still far too small to resolve the
substructure crisis.  It seems doubtful at present that new searches
will add the hundreds of objects required to solve this problem -- if
the missing dark matter halos contain luminous matter as well.

Another promising avenue is the search for stellar streams in the
halos of massive galaxies (e.g., \cite{Bullock01, Newberg02,
Newberg03, Yanny03, Martin04, Rocha04, Juric05, Pena05, Wyse06,
Belokurov06b}), which may ultimately permit us to constrain the number
of past accretion events, esp.\ once full phase-space information is
added -- one of the objectives of ESA's Gaia satellite mission
(\cite{Perryman01}).

Interestingly, the companions of the Milky Way do not seem to show a
random distribution. Instead, they appear to be arranged along one or
two polar great planes (e.g., \cite{Kunkel79, Lynden82, Majewski94,
Fusi95, Kroupa05}).  An investigation of the satellite distribution
around M31 revealed that most of its early-type companions lie along
one single, almost polar great plane with high statistical
significance (\cite{GreKolBra99, KG06}, see also \cite{McConn06}).  As
shown by \cite{KG06}, this plane does not coincide with the great
stellar stream around M31.  It is unclear whether these apparent
satellite anisotropies have a deeper physical significance.  If the
satellites orbit within the planes -- a requirement if the planes have
any physical reality -- then the satellites may have formed following
interactions or the break-up of a more massive progenitor (see, e.g.,
\cite{Kunkel79, Lynden82, Sawa05, Bournaud06}).  Another possibility
is that the planes are indicators of a prolate dark halo of the Milky
Way and M31 (e.g., \cite{Hartwick00, Navarro04, Zentner05}).  Or we
may be seeing the left-over effects of matter accretion along
large-scale dark matter filaments (see \cite{Knebe04, Zentner05,
Libeskind05} for a more detailed discussion).  \cite{KG06} found that
the M31 polar plane includes also M33 and points in the direction of
the M81 group.  -- In the absence of actual orbits it remains
difficult to assess the meaning of the observed polar alignments, if
any.  For the M31 subsystem the planar alignment comprises only a
subsample of the total number of its satellites (\cite{KG06}), which
may lend support to the suggestion by \cite{Zentner05} that such
distributions may also originate from random samples.  Unfortunately,
the distance uncertainties for galaxies in nearby groups are still too
large, preventing us from conducting a similar study there.

\subsection{A few remarks about star formation histories}

The star formation histories of the galaxies of the Local Group have
been reviewed fairly frequently (e.g., \cite{vdB00, Grebel00,
Grebel05}).  Here we will only summarize some interesting
characteristics that have emerged in recent years.  

Old populations are ubiquitous in the galaxies of the Local Group, but
their fractions vary (\cite{Grebel00, GreGal04}).  For instances,
horizontal branches are an unmistakable tracer of old populations and
have been identified in all Local Group galaxies with sufficiently
deep photometry.  Here ``old'' refers to old Population II stars with
ages $> 10$ Gyr.  We have no evidence for the existence of possible
Population III stars in external galaxies.  For galaxies with deep
main-sequence photometry (mainly Galactic populations and satellites
of the Milky Way) there is evidence for a common epoch of early
(Population II) star formation.  Within the currently available
accuracy of the age dating techniques ($<$ 1 Gyr), the oldest
detectable populations in the Milky Way and its surroundings are
coeval (\cite{GreGal04}).  Moreover, there is no evidence for the
suppression of star formation in the low-mass galaxies of the Local
Group after the end of reionization (\cite{GreGal04}), contrasting
predictions of certain cosmological models that suggest that low-mass
halos experience complete photo-evaporation (\cite{Barkana99,
Ferrara00, Susa04}).

All Local Group galaxies show evidence for extended star formation
histories, but no two galaxies share the same star formation history,
not even within the same morphological type (e.g., \cite{Grebel97}).
The spirals, irregulars, and dIrrs all show evidence for active star
formation over a Hubble time.  In dwarf galaxies star formation
proceeded largely continuously with amplitude variations.  Strongly
episodic star formation with long pauses in between is only seen in
the Carina dSph (e.g., \cite{Monelli03}).  Even dSph galaxies with
seemingly entirely old populations show large abundance spreads of
more than 1 dex in [Fe/H] (e.g., \cite{Shetrone01}), which require
star formation and enrichment over several Gyr (\cite{Ikuta02,
Marcolini06, Fenner06}).  In contrast to many dIrrs, which are able to
continue to form stars for another Hubble time (\cite{Hunter97}),
dSphs do not show any ongoing star formation activity and appear to be
devoid of gas (\cite{Gallagher03} and references therein).  This is
unexpected, since even simple mass loss from red giants should lead to
detectable amounts of gas (e.g., \cite{Young00}).  The Milky Way is
surrounded by several dSphs with substantial intermediate-age (2 to 10
Gyr) star formation activity.  One of the Galactic companions, the
Fornax dSph, even ended its star formation activity only as recently
as about 200 Myr ago (\cite{GreSte99}).  For a more detailed
discussion of the gas loss problem see \cite{GGH03}.    
 
Population gradients are obvious in spiral galaxies, but also many
dwarf galaxies show spatial variations in their star formation
histories (e.g., \cite{Grebel99, Grebel00}).  Essentially all galaxies
have extended ``halos'' of old Population II stars (e.g.,
\cite{Minniti99}).  DIrrs tend to also show extended intermediate-age
populations as traced by, e.g., carbon stars (e.g., \cite{Letarte02}).
Generally, the density distributions of different populations in
irregular galaxies become increasingly more regular and extended with
increasing age (e.g., \cite{Cioni00, Zaritsky00}), whereas the many
scattered young star-forming regions are responsible for the irregular
appearance.  Large star-forming regions may remain active for several
10 to 100 Myr, revealing a complex substructure in ages (e.g.,
\cite{Dohm97, GreBra98, Walborn99, GreChu00, Dieball00, Dieball02}).  

DE and dSph galaxies have usually experienced continuous star
formation with decreasing intensity.  Star formation tends to be
longest-lasting in the centers of these galaxies, and in a number of
cases radial age (and possibly metallicity) gradients are observed
(see \cite{Grebel97, Stetson98, Hurley99, DaCosta00, Sarajedini02,
Tolstoy04, Koch06} for individual cases and \cite{Harbeck01} for a
comprehensive study).  Subpopulations may also be asymmetrically
distributed (e.g., \cite{Stetson98}).  Overall, the Galactic dSph
companions show a trend for increased intermediate-age population
fractions with increasing Galactocentric distance, possibly indicating
that star-forming material might have been removed earlier from closer
companions (\cite{vdB94, Grebel97}).  However, neither the fairly
isolated Local Group dSphs Cetus and Tucana fit this pattern, nor do
the M31 low-surface-brightness companions (\cite{GGH06}).   

In the past star formation histories of galaxies with resolved stellar
populations were primarily based on photometry, and on the modelling
of the observed color-magnitude diagrams (as pioneered by
\cite{Tosi91, Greggio93}).  Limitations of modelling techniques are
discussed in \cite{Gallart05}.  Additional complications may arise
from rotation, multiplicity, unrecognized extinction effects, and
transformation problems (e.g., \cite{GR95, Roberts95, GRB96,
Zaritsky02, Koch04, Girardi04, Zaritsky04}).  While in many cases no
other information but photometric data are available, information on
special types of stars can provide important additional clues (e.g.,
\cite{Grebel97, Grebel99, Borissova00, Battinelli00, Dolphin02,
Harbeck04, Harbeck05}).  Special types of stars can be identified
photometrically if, e.g., they exhibit unique spectral features that
can be traced by special filter combinations (e.g., \cite{Cook86,
Cook89, GRdB92, Grebel97b, Majewski00, Nikolaev00, Keller01, Cioni03,
Helmi03, Harbeck04, Kerschbaum04, Kniazev06}) or if they are variable
(e.g., \cite{Mateo95, Kovacs95, Mateo98, Kovacs01, Dolphin02,
Pritzl05b}).  

Photometric systems with improved metallicity sensitivity have been
employed to break the photometric age-metallicity degeneracy (e.g.,
Washington and Str\"omgren photometry, \cite{Geisler86, Geisler91,
GR92, GRvdR94, Hilker00, Cole00}).  However, this degeneracy can best
be addressed spectroscopically. This is now a realistic prospect for
nearby galaxies (e.g., \cite{Cote99, Cole00, Guha00, Shetrone01,
Pont04, Tolstoy04, Koch06, Koch06b}), largely thanks to the routine
availability of optical 8 to 10-m telescopes.

\subsection{A few remarks on abundances}

While the absorption-line measurements for stellar abundances tend to
require large telescopes and long integration times, emission lines
can relatively easily be measured with medium-sized telescopes.  In
galaxies with active star formation the metallicity of young
populations can be determined by measuring the emission lines of
H\,{\sc ii} regions.  This has not only been done for gas-rich
galaxies such as dIrrs in the Local Group and its surroundings (e.g.,
\cite{Skillman89, Saviane02, Lee03, Kniazev05}, but is routinely
carried out also for much more distant galaxies with sufficiently
strong emission lines (e.g., \cite{Kniazev04a}).  These measurements
yield the present-day abundances, and individual element abundance
ratios give us information about the modes of star formation.  

Also nebular abundances derived from planetary nebulae can be obtained
with only a modest investment of telescope time, yielding metallicity
estimates for mainly intermediate-age populations (e.g.,
\cite{Richer98, Jacoby99}), which can be age-dated within certain
limits (\cite{Kniazev05, Buzzoni06}).  Combining ages with the results
from nebular abundances as well as stellar abundance measurements
supports the existence of an age-metallicity relation, i.e.,
increasing enrichment with decreasing age (e.g., \cite{DaCosta98,
Tolstoy03, Grebel04, Kniazev05, Koch06}).  The comparison with
chemical evolution models can then provide fairly detailed information
about the enrichment history as well as the relative importance of
closed-box evolution vs.\ outflows and galactic winds (e.g.,
\cite{Mighell98, DaCosta98, Lanfranchi03, Lanfranchi04, Hensler04,
Koch06}).

For a given age, a given dIrr galaxy is usually assumed to be
chemically well-mixed and homogeneous.  However, there are indications
in a few dIrrs for a metallicity spread in populations of similar ages
based on age-datable star clusters (\cite{DaCosta02}) or on differing
nebular abundances in H\,{\sc ii} regions (\cite{Kniazev05}).  For
dSphs, we do not yet have data that would permit us to quantify
possible abundance spreads within populations of the same age, but as
noted earlier dSphs exhibit large metallicity spreads overall, even in
galaxies dominated entirely by old populations (e.g.,
\cite{Shetrone01}). 

Interestingly, high-resolution studies of individual red giants in
nearby dwarf galaxies indicate that at a given [Fe/H] their
[$\alpha$/Fe] ratios are on average lower by $\sim 0.2$ dex than those
of equally metal-poor stars in the Galactic halo (e.g., \cite{Hill00,
Shetrone01, Shetrone03, Boni04, Geisler05, Pritzl05}).  This makes
these {\em present-day} dwarf galaxies -- dIrrs and dSphs alike --
unlikely major contributors of the build-up of the Galactic halo and
hence unlikely survivors of a once much more numerous population of
primary Galactic building blocks.  However, at low [Fe/H] there is
consistency with the abundance ratios observed in the Galactic halo,
leaving very early accretion as a plausible option (\cite{GGH03,
Font06}).

On global scales, galaxies tend to follow a luminosity--metallicity
relation in the sense that more luminous galaxies are more metal-rich.
This trend is seen for all Local Group galaxies, but the
metallicity--luminosity relations of dSphs and dIrrs are offset from
each other (\cite{Richer98}), an offset that persists even when
comparing the same metallicity tracers and the same populations
(\cite{GGH03}): DSphs have higher mean stellar metallicities at a
given optical luminosity, which may imply more rapid star formation
and enrichment at early times as compared to dIrrs (\cite{GGH03}).
This suggests that the old populations in dSphs and dIrrs are
intrinsically different.  Transitions from gas-rich to gas-poor dwarfs
seem plausible only for present-day dIrr/dSphs to present-day dSphs.

\subsection{A few remarks on kinematics and dark matter}

The internal kinematics of galaxies are not only a valuable tool to
differentiate their components, populations, and evolutionary
histories, but also provide information about galaxy masses.  For the
large spiral galaxies, a variety of approaches have been used to
constrain their masses, including stellar and gas kinematics, the
kinematics of star clusters, and of satellites.  These data seem to
suggest that the Milky Way and M31 may be of similar mass
(\cite{Kochanek96, Evans00, Carignan06}).  This is somewhat 
unexpected considering the larger luminosity, larger bulge, larger
number of globular clusters, and larger physical size of M31.   

While the disks of spiral galaxies exhibit differential rotation, the
more massive dIrrs show solid-body rotation.  Low-mass dIrrs,
dIrr/dSph and dSph galaxies are dominated by random motions and do not
appear to be supported by rotation.  DSphs may contain large amounts
of dark matter (e.g., \cite{Gallagher94}).  This is inferred from the
high velocity dispersion and the resulting high mass-to-light ratios
derived under the assumption of virial equilibrium (\cite{Mateo97}).
Indirectly, a high dark matter content is also supported by the
smooth, symmetrical morphology of some nearby dSphs (\cite{Odenk01})
and by the observed lack of a significant depth extent
(\cite{Klessen03}).  The radial velocity dispersion profiles of dSphs
tend to be flat and fall off large radii (\cite{Wilkinson04,
Wilkinson06}).  While detailled modelling is still in progress, the
current data favor that dSphs share a common halo mass scale of about
$4\cdot 10^7$ M$_{\odot}$, are dark-matter dominated, and have cored
mass distributions (\cite{Wilkinson06b}).

Evidence for ongoing accretion events in the Local Group have been
mentioned already in Section 2.1.  Stellar streams may also be
contributed by disrupted globular clusters (e.g., \cite{Odenk01b,
Odenk03, Belokurov06c, Grillmair06}) and can provide valuable
information about the shape of the massive galaxy's halo (e.g.,
\cite{Murali99, Bullock01, Dehnen04, Pena05, Johnston05, Majewski06}).
The results can be compared to the halo shape derived from other
indicators such as the halo globular cluster mass density profile,
which seems to be partly of primordial origin (\cite{Parmentier05}).
--- A well-known example of ongoing interactions is the triple system
of the Magellanic Clouds and the Milky Way with the gaseous Magellanic
Bridge and Magellanic Stream as interaction signatures.  Tidal
interactions are also apparent in the S-shaped surface density profile
of the Galactic dSph Ursa Minor (\cite{Palma03}) and in the twisted
isophotes of the M31 dE companions M32 and NGC\,205 (\cite{Choi02}).
These galaxies and other dSphs are likely to be accreted eventually.
Moreover, dwarf-dwarf interactions and interactions with gas clouds
may play an important role (e.g., \cite{Wilcots98, deBlok00,
Coleman05}). The Local Group's wealth in low-mass, gas-deficient
early-type dwarfs as well as the radius-morphology relation may both
be indicative of the environmental impact on galaxy evolution.

\section{Other nearby groups and Local Group analogs}
\label{sec:2}

In our immediate cosmic neighborhood we find poor, loose groups and
``clouds'' or filaments.  The spatial and density distribution of
nearby galaxies is beautifully illustrated in \cite{TFatlas87}.  A few
years ago we initiated a project with the Hubble Space Telescope and
ground-based telescopes to study the properties of the groups and
clouds within a $\sim 5$ Mpc around the Local Group, i.e., in the
Local Volume (\cite{Gsnap00}).  This project led to an improved galaxy
census and resulted in knowledge of the approximate distances,
luminosities, luminous stellar content, and approximate metallicities
of a large number of nearby galaxies (\cite{Ksnap99, KKSG00, Ksnap00,
KM8100, Dolphin01, KM8101, KKR2501, Kara02b, Maka02, Sarajedini02,
Kara02c, MakaM8102, Kara03b, Kara03c, Kara03d, Lee03, Kniazev03,
Maka05, Kniazev05}; see also \cite{Cote97, Caldwell98, Jerjen00,
Barazza01, Parodi02, Rejkuba06}).  Moreover, kinematic properties and
global masses were derived, improving the characterization of these
nearby groups (see also \cite{Cote00, Kara02a, Kara03a, Tully05,
Kara05}).  Many of the results presented in the following stem from
this continuing project.  

\cite{Plionis04} find that galaxy groups are considerably more
elongated than galaxy clusters, a trend that becomes most pronounced
in very poor groups of galaxies.  They suggest that the poorest groups
of galaxies are still in the process of being assembled through galaxy
infall, a scenario that is supported by the observed properties of
nearby, extended galaxy ``clouds'' like Sculptor and CVn I
(\cite{Jerjen98, Kara03b, Kara03c}):  These elongated clouds show
several subclumps and are not yet in dynamical equilibrium, since
their crossing times are of the order of half a Hubble time.  They
contain mainly early-type dwarfs, and their luminosity functions show
a lack of very low-luminosity dSph galaxies.  CVn I may be even less
evolved than Scl.  CVn I, Scl, and the Local Group form a 10 Mpc
filament that appears to be driven by the free Hubble flow
(\cite{Kara03c}).   

In contrast to the clouds, the nearby, poor groups are Local Group
analogs in many respects.  They tend to be dominated by two massive
galaxies: In the M81 group, the dominant galaxies are the spirals M81
and NGC\,2403 (\cite{Kara02b}).  A prominent interaction between M81,
and the irregular galaxies M82 and NGC 3077 is currently in progress
(e.g., \cite{Yun94}), giving rise to impressive starburst phenomena in
M82 (e.g., \cite{OConnell95}) and possibly to the formation of tidal
dwarfs from material torn out during past close encounters (e.g.,
\cite{MakaM8102}).  In the Centaurus A group the spiral M83 and the
peculiar elliptical Cen A (NGC 5128) dominate the mass distribution
(\cite{Kara02c}).  NGC\,5128 seems to have experienced various
accretion events in the past (e.g., \cite{Malin83, Mirabel99}).  In
the IC\,342/Maffei group the main subgroups are centered around the
spiral IC\,342 and the elliptical Maffei I (\cite{Kara03d}).  In this
group the faint galaxy census is still highly incomplete due to the
high foreground extinction (e.g., \cite{Buta99}).  In any case, all of
these nearby groups reveal a ``binary'' substructure.

Just like the Local Group, the nearby groups also exhibit an increased
frequency of early-type dwarfs, particularly of dSphs, and a
comparable degree of morphological segregation (\cite{Grebel04,
Grebel05}).  The luminosity functions show the familiar rise at the
faint end (\cite{Kara02c, Kara03b}).  The differences in the
fractional dwarf galaxy type distribution between clouds and Local
Group analogs supports the idea of morphological transformations
induced by denser environments.  In accordance with this picture,
transition-type dIrr/dSph galaxies and dIrrs tend to be found at
larger distances from massive galaxies (e.g., \cite{GGH03}), and dwarf
S0 galaxies are located in the outskirts of groups (and clusters;
e.g., \cite{Beaulieu06, Lisker06}).  \cite{Mahdavi99} note that
star-forming galaxies are more likely on radial orbits and may be
falling in for the first time, whereas the orbits of galaxies
dominated by older populations are more consistent with isotropy.

In the M81 group, the M81 subgroup and the NGC\,2403 subgroup approach
each other, similar to what is seen in the Local Group between M31 and
the Milky Way.  A third, smaller subgroup around the less luminous
spiral NGC 4236 in the M81 group appears to be currently receding from
the other two subgroups (\cite{Kara02b}) and may actually even
constitute a small group outside of the M81 group.  In the Cen\,A
group the two dominant subgroups also appear to be moving away from
each other (\cite{Kara02c}).  We have insufficient data on the other
groups. -- Where subgroups approach each other the final result is 
expected to be  one single large elliptical galaxy (see also
\cite{Forbes00}) and hence a fossil group (\cite{Jones03}).  

The estimated crossing times in the nearby, poor groups range from 1.8
to 5.9 Gyr with a median around $\sim 2.3$ Gyr (\cite{Kara05}),
suggesting that they are closer to reaching dynamical equilibrium than
the unevolved clouds, although they are far from ``fossilization''.
The radii of the zero-velocity surfaces (beyond which galaxies are no
longer bound) of the nearby groups and of the Local Group are of the
order of 1 Mpc, and the total masses of the groups are approximately a
few $10^{12}$ M$_{\odot}$ (\cite{Kara02b, Kara02c, Kara03d}).  Within
groups, the mass is closely correlated with the luminous matter and
concentrated at the location of the massive galaxies (in agreement
also with the findings by \cite{McKay02, Prada03}). 

On larger scales, the nearby groups and clouds nicely delineate
the local large-scale structure (\cite{TFatlas87}).  The centers of
the groups lie within a narrow layer with a thickness of only $\pm
0.33$ Mpc (\cite{Kara05}).  The centroids of the groups show a small
velocity dispersion with respect to the Hubble flow.  Overall the
local Hubble flow is remarkably cold (\cite{Kara02a, Kara03a}),
indicating a low local matter density.  \cite{Maccio05} suggest that
the local Hubble flow is best fit by $\Lambda$CDM models with
signatures of the impact of dark energy already becoming noticeable at
distances $> 7$ Mpc.  

\section{Summary}

In the Local Group, old populations are ubiquitous in all galaxies,
but their fractions vary.  There appears to have been a common epoch
of early (Population II) star formation.  All galaxies show evidence
for extended star formation episodes, but no two galaxies share the
same detailed star formation history.  There is no obvious cessation
of star-formation activity after re-ionization in low-mass galaxies.
The apparent correlation between increasing intermediate-age
population fraction in dSph galaxies with increasing distance from the
Milky Way is not seen in the M31 dSph companions.  The morphological
segregation and H\,{\sc i} mass -- distance correlation in the Local
Group and other nearby groups hint at the importance of environment
and interactions.  This is also indicated by the observed ongoing
interactions and by the increased fraction of low-luminosity,
early-type dwarfs in groups as compared to loose, unvirialized
``clouds'' of galaxies.  However, there is an offset in the
metallicity-luminosity relation even for old populations that shows
dSphs to be too metal-rich for their luminosity in comparison to
dIrrs, making a simple transformation from dIrrs to dSphs unlikely.
The origin and nature of dSphs remains a puzzle.  Also, the meaning
(if any) of  the seemingly anisotropic distribution of the early-type
companions of the Milky Way and of M31 along a polar great plane
remains unclear as long as accurate orbits are lacking.   

Nearby groups in the Local Volume are Local Group analogs in many
respects.  Their luminosity functions, galaxy content, and fractional
type distribution are reminiscent of the Local Group. They show
morphological segregation.  Just like the Local Group, they are
typically dominated by two luminous galaxies.  The crossing times,
group masses and radii resemble those of the Local Group.  These poor
groups may ultimately evolve into fossil groups once their subgroups
merge.  This stands in contrast to nearby galaxy ``clouds'', which
have long crossing times, are extended, have few early-type dwarfs and
are dominated by gas-rich late-type galaxies.  These clouds do not
show a turn-up at the faint end of the galaxy luminosity function.
Perhaps they are groups in formation. --- The local Hubble flow is
very quiet. 

\input{grebelref}



\printindex
\end{document}

%% file: grebelref.tex
%
%

%
%